\documentclass[journal]{IEEEtran}
\ifCLASSINFOpdf
\else
\fi

	\usepackage{graphicx}
	\usepackage{url}

	\usepackage{draftwatermark}
	\SetWatermarkText{DRAFT}
	\SetWatermarkScale{1.0}  
	\SetWatermarkColor[gray]{0.88}  
	\usepackage{amsmath}
	
\begin{document}



		%
		\title{Radar and Acoustic Sensor Fusion using a Transformer Encoder for Robust Drone Detection and Classification}

		%
		\author{\IEEEauthorblockN{Gevindu~Ganganath$^{*}$, Pasindu~Sankalpa$^{*}$,
				Samal~Punsara$^{*}$,~Demitha~Pasindu$^{*}$, Chamira~U.~S.~Edussooriya$^{*}$$^{1}$,~Ranga Rodrigo$^{*}$$^{2}$,~and Udaya~S.~K.~P.~Miriya~Thanthrige$^{*}$$^{1}$}\\ 
			\IEEEauthorblockA{$^{*}$Department of Electronic and Telecommunication Engineering,
				University of Moratuwa, Sri Lanka.\\
				$^{1}$Member, IEEE,~
				$^{2}$Senior Member, IEEE.}
			
			\thanks{{Corresponding author: Udaya SKPM Thanthrige (e-mail: sampathk@uom.lk).~\\ © 2026 IEEE. Personal use of this material is permitted. Permission from IEEE must be obtained for all other uses, in any current or future media, including reprinting/republishing this material for advertising or promotional purposes, creating new collective works, for resale or redistribution to servers or lists, or reuse of any copyrighted component of this work in other works. Accepted at IEEE 12$^{\text{th}}$~Moratuwa Engineering Research Conference 2026 (MERCon 2026)}.\protect
			}
		}


		\maketitle
		
		\begin{abstract}
			The use of drones in a wide range of applications is steadily increasing.  However, this has also raised critical security concerns such as unauthorized drone intrusions into restricted zones. 
			Therefore, robust and accurate drone detection and classification mechanisms are required despite significant challenges due to small size of drones, low-altitude flight, and environmental noise. In this letter, we propose a multi-modal approach combining radar and acoustic sensing for detecting and classifying drones. We employ radar due to its long-range capabilities, and robustness to different weather conditions. We utilize raw acoustic signals without converting them to other domains such as spectrograms or Mel-frequency cepstral coefficients. This enables us to use fewer number of parameters compared to the state-of-the-art approaches. Furthermore, we explore the effectiveness of the transformer encoder architecture in fusing these sensors. Experimental results obtained in outdoor settings verify the superior performance of the proposed approach compared to the  state-of-the-art methods.
			
		\end{abstract}
		
		
		\section{Introduction}
		
		Unmanned aerial vehicles (UAVs), commonly known as drones, have significantly evolved, enabling applications in agriculture, cinematography, infrastructure inspection, and security. However, these advancements have also raised critical security concerns~\cite{Khan2022}. Unauthorized drone intrusions into restricted zones, including airports and military bases, pose threats to public safety and national security. Drones have also been exploited for smuggling contraband into secure facilities and for illicit surveillance, raising privacy concerns~\cite{Srirangam2023}. Furthermore, recently drones played a significant role in modern warfare, emphasizing the urgency of robust detection and mitigation systems~\cite{10863087, 10713174,10685581}.
		
		Various techniques have been used for drone/UAVs detection and classification, and radar based approaches are widely been employed due to their robustness in various weather conditions and long range detection capabilities~\cite{9444586}.
		Traditional drone detection methods, including radar, optical, and acoustic sensors, face notable challenges. Optical detection, although cost-effective, suffers from limited range and sensitivity to environmental factors such as fog and lighting conditions \cite{jung2018avss}. These methods leverage visual processing techniques, such as the histogram of oriented gradients and deep learning-based classifiers, for drone identification \cite{jung2018avss}. Acoustic-based detection relies on unique sound signatures and employs machine learning and spectrogram analysis~\cite{kolamunna2021droneprint}, but is prone to interference from background noise and adverse weather conditions~\cite{kolamunna2021droneprint}. Radio frequency (RF)-based detection exploits wireless transmissions to classify drone types and flight modes~\cite{al2019rf}.~However, its effectiveness diminishes with encrypted or frequency-hopping drones and fully autonomous drones. Radar, which does not depend on drone communication signals, remains a crucial technology but struggles with false positives from low-radar cross section (RCS) objects such as birds \cite{liu2017digital}. 
		
		Due to these limitations, recent research has focused on the multi-modal fusion of sensors, integrating complementary sensing modalities to improve detection robustness and precision~\cite{lee202310cnn, svanstrom2021real, 9944867}. To this end, deep learning-based methods have shown improvements in drone detection in real-time, yet transformer-based fusion techniques remain under explored~\cite{vaswani2017attention}. 
		These fusion techniques have demonstrated successful applications in diverse domains such as autonomous driving \cite{bai2022transfusion}, industrial sensor fusion \cite{ nath2021structural}, target tracking \cite{wei2024transformer}, odometry estimation \cite{sun2023transfusionodom}, and event detection \cite{yasuda2022multi}. By leveraging a transformer‐based fusion backbone, above works achieve effective cross-modal attention, exhibit robustness to missing or corrupted channels through learned attention masks, and offer inherent scalability and transfer-learning capabilities, enabling seamless integration of additional modalities. The contributions of this letter are summarized as follows. 
		\begin{enumerate}
			\item We propose drone detection and classification based on radar and acoustic in outdoor settings. To the best of our knowledge, there are not many studies which present experimental results in outdoor.  We employ radar due to its long-range capabilities, robustness to different weather conditions, and ability to track fast-moving objects, whereas the fusion of radar and acoustic signal improve the performance.
			\item We validate our proposed approach in outdoor settings using frequency-modulated continuous wave (FMCW) radar and a directional microphone. Here, we demonstrate state-of-the-art performance with potential for real-world deployment. In addition, we publish the collected dataset\footnote{\tiny{\url{https://drive.google.com/drive/folders/1_ApoTYP2nhK8c0PnDvNYvvwdUy6HXK3g}}} to facilitate further research in this domain. 
			
			\item Our proposed approach is based on a multi-modal deep learning model that integrates radar and acoustic data. Our approach utilizes the raw acoustic signal without converting any domain such as Mel. Thus, this eliminates the need for additional preprocessing
			step and this enables us to use fewer number of parameters compared to state-of-the-art.

		\end{enumerate}
		
		\section{Measurement Setup and Data Collection}
		
\begin{figure}[h]
			\centering
			\includegraphics[ width=0.98\linewidth]{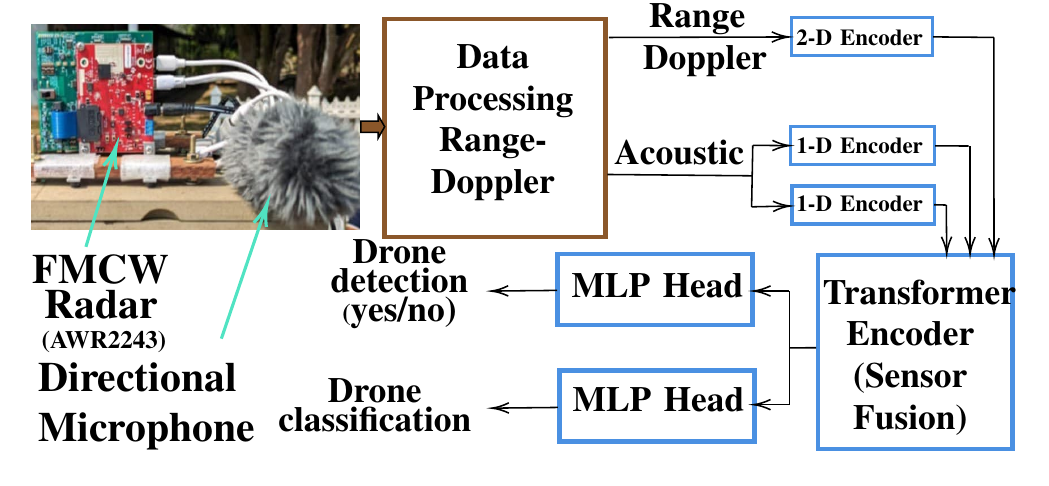}
			\caption{Measurement setup with a FMCW radar and a directional microphone followed by the deep learning model.}\label{fig:hsetup}
		\end{figure}

		In this work, frequency-modulated continuous wave (FMCW) radar (Texas Instrument AWR$2243$ Boost~\cite{AWR2243}) and a directional microphone is used for data collection. The measurement setup and the data processing chain are shown in Fig.~\ref{fig:hsetup}. The specifications of the FMCW radar module is presented in Table~\ref{Radar Parameters}. The radar module offers high-resolution sensing which is ideal for drone detection. For the acoustic data collection, we employed a directional microphone with a $16$~kHz sampling rate. 
		
		\begin{table}[ht!]
			\caption{Specifications employed with the AWR2243 radar.}
			\label{Radar Parameters}
			\centering
			\begin{tabular}{lclc}
				\hline
				\hline
				Parameter & Value & Parameter & Value\\
				\hline
				Start Frequency (GHz)     & 77   &   Samples per chirp         & 256        \\
				Chirps per frame          & 128  &    Sampling rate (Msps)      & 10      \\
				Frame periodicity (ms)  & 40   &      Bandwidth (GHz)           & 0.76    \\
				Slope (MHz/$\mu$s)        & 29.98 &    Range resolution (m)      & 0.19   \\
				Velocity resolution (m/s) & 0.15  &    No. of TX / RX antennas& 1 / 1    \\       
				\hline
			\end{tabular}
		\end{table}
		
		The FMCW radar transmits a continuously varying frequency and measuring the frequency difference between the transmitted and received signals to estimate the range and velocity of the target. Here, we follow the standard procedure to estimate the range-Doppler map of the target, that is, mixing the received signal with the transmitted signal, followed by applying fast Fourier transform to extract range and Doppler information of the target~\cite{9444586,10965358}. To reduce the impact of ground clutter and noise, zero-Doppler filter and the constant false alarm rate (CFAR) algorithm are used.

		For acoustic data collection, a sampling rate of $16$~kHz is used in stereo mode (two channels). During processing, audio signals were converted to mono (one channel) and normalized to compensate for amplitude variations due to flight distance and microphone characteristics. Next, data collection process is discussed.
		
		We collected data for five cases: no-drone, DJI Matrice $300$ RTK, Mavic $2$ Enterprise Dual, DJI Phantom $4$ Pro Plus and DJI Phantom $4$ Pro V$2.0.$ For each drone class, we collected $2048$ samples and for non-drone class we collected $2432$ samples. For each drone type, we considered several scenarios such as the drone flying towards the radar in several heights while facing different sides of the drone, flying in cross patterns in several heights, and random movements. For radar, we collected binary files for each of the flight scenarios. Each binary file has $256$ frames, each frame has $128$ chirps and each chirp has $256$ samples. For acoustic data, we collected roughly $12$ seconds long .wav files.
		
		\section{Proposed Deep Learning Model}
		In this section, the proposed deep learning model for drone detection and classification is presented. The overall model architecture, shown in Fig.~\ref{fig:hsetup}, has three main stages: the encoder stack, transformer-based fusion and multi-layer perceptron (MLP) heads. Here, range-Doppler and acoustic data are fed to the proposed transformer-based deep learning model as different modalities. One-dimensional or two-dimensional counterparts of the encoder are used depending on the input dimensionality.
		
		The proposed encoders employ convolution for feature extraction motivated from the work done by Hunje Lee et al.~\cite{lee202310cnn}. Residual connections are used in order to avoid vanishing gradient problem due to higher depths~\cite{he2016deep}. The $K$ number of squeeze and excitation blocks are employed to activate between extracted channels~\cite{hu2018squeeze}. In order to perform an end-to-end audio classification, we followed the method proposed by~\cite{kumar2020end}, which helps to extract features from different frequency ranges. We use only two encoders and perform an element-wise addition to combine the extracted features. The selection of kernel size is important as it significantly affects performance. Thus, we experiment with various kernel sizes and selected the configuration based on classification and detection accuracy. For acoustic data, kernel sizes of $7$ and $107$ are used as the small and large kernel sizes, respectively, for one-dimensional encoders. An additional down-sampling is performed with a convolution of kernel size $15$ due to the large input size. 
		$K$ is set to $5$ for all one-dimensional encoders. A two-dimensional encoder is used for the range-Doppler data. No down-sampling is performed. $K$ is set to $4$ with kernel sizes as $3,~3,~5$ and $7$, respectively.
		
		\subsection{Sensor Fusion}
		
		Rather than using traditional machine learning algorithms, which has demonstrated low performance in classification, we use a more sophisticated fusion mechanism based on self-attention. Therefore, we utilize the transformer encoder~\cite{vaswani2017attention} initialized to have eight heads, one layer with the embedding dimension $128$ as our fusion method. 
		
		In literature, there are attempts such as~\cite{lee202310cnn} for drone detection and classification using two stage architectures. Here, they have to train two separate stages of model with several encoders which requires a much effort and it also adds an extra computational cost in inference time which can affect the real-time performance. To overcome these issues, we propose to use two separate MLP heads, one for detection and one for classification. 
		
		To facilitate the joint training of the detection and classification heads, a custom loss function is defined as 
		\begin{equation}
			\label{loss}
			\text{Loss} = \frac{1}{N} \sum_{n=1}^{N} X_{\text{ent}}(\hat{y}_{\text{det}, i} , y_{\text{det}, i}) + \lambda~y_{\text{det}, i}~X_{\text{ent}}(\hat{y}_{\text{cls}, i} , y_{\text{cls}, i}),
		\end{equation}
		where the first term, $X_{\text{ent}}(\hat{y}_{\text{det}, i} , y_{\text{det}, i})$ represents the cross-entropy between the predicted detection output $\hat{y}_{\text{det}, i}$ and the ground truth detection label $y_{\text{det}, i}$, and $y_{\text{det}, i} \in \{0, 1\}$ indicates the absence or presence of drone, respectively. The second term corresponds to the classification loss, defined as the cross-entropy between the predicted class label $\hat{y}_{\text{cls}, i}$ and the true class label $y_{\text{cls}, i}$. This term is multiplied by $y_{\text{det}, i}$, ensuring that the classification loss is only considered when an object is detected (i.e. $y_{\text{det}, i} = 1$). Furthermore, $\lambda$ is a regularization parameter that controls the relative importance of the classification loss, which helps to stabilize training. The summation is performed over a mini-batch of size $N$, and the total loss is normalized by $N$ to obtain the average loss across the batch. This formulation effectively suppresses unnecessary classification loss in the absence of a detected object while allowing joint optimization of both detection, and classification objectives.
		
		
		
		
		\subsection{Model Training} \label{model training section}
		
		In this subsection, we discuss how we tackle the problem of data imbalance when training both MLP heads jointly. At the end, we present model hyper-parameters. An important factor to be aware of before starting the training is the number of samples for each class in the dataset. Output classes are drone and non-drone for the detection head, whereas they are non-drone and four drone classes for the classification head. Initially, each of the five classes in the dataset has roughly the same number of samples. This is good for the training of the classification head, but for the training of the detection head, since all drone classes are combined into a single class, a class imbalance is occurred. This is a conflict between two heads because the dataset should be class balanced with respect to both heads in order to perform proper joint training. As a solution, the non-drone class is up-sampled to match the number of samples in the combined drone class. At first sight, this seems to cause a class imbalance at the classification head. But there is no such effect due to the non-drone class canceling in the loss function because, 
		\begin{equation}
			\label{case}
			X_{\text{ent}}(\hat{y}_{\text{cls}} , y_{\text{cls}}) = \begin{cases}
				X_{\text{ent}}(\hat{y}_{\text{cls}} , y_{\text{cls}}) & y_{\text{det}} = 1\\
				0 & y_{\text{det}} = 0.
			\end{cases}
		\end{equation}
		Effectively, the classification head trains only for the four drone classes which are class balanced.
		
		Our final fusion model has $68.5$~G floating point operations and $15$~M parameters. The model is trained on an NVIDIA GeForce RTX $3090$ graphics processing unit for $60$ epochs with a batch size of $64$. We employ Adam optimizer and the initial learning rate is set to $5\times10^{-5}$ whereas the weight decay us set to $0.4$. All the dropouts are set to $0.4$. Early stopping is used to avoid over-fitting.
		
		\section{Experimental Results}
		
		We now present experimental results to verify the effectiveness of the proposed approach. To this end, we first compare performance of our model with several existing methods. We evaluate the system described in~\cite{lee202310cnn} on our dataset and compare with our acoustic and range-Doppler fusion. Table \ref{table:comp} shows the comparison. Accordingly, the proposed deep learning method \emph{outperforms~\cite{lee202310cnn} significantly}. 
		\begin{table}[t!]
			\caption{Performance of the proposed model and the work done by Hunje Lee et al.~\cite{lee202310cnn}.}
			\label{table:comp}
			\centering 
			\resizebox{\columnwidth}{!}{
				\begin{tabular}{lcccc}
					\hline
					\hline
					\textbf{Architecture} & \multicolumn{2}{c}{\textbf{Detection}} & \multicolumn{2}{c}{\textbf{Classification}} \\
					& accuracy (\%) & F1-score & accuracy (\%) & F1-score\\
					\hline
					Hunje Lee et al. (DenseNet)~\cite{lee202310cnn}	 & $75.00$ \%	& $0.7494$ & $68.34$ \% & $0.6733$ \\
					Hunje Lee et al. (ResNet)~\cite{lee202310cnn} & $75.00$ \% & $0.7491$ & $64.09$ \% & $0.6179$ \\
					\textbf{Ours (Acoustic + Range-Doppler)} & \textbf{$100.00$} \% & \textbf{$1.0000$} & \textbf{$100.00$} \% & \textbf{$1.0000$} \\
					\hline
			\end{tabular}}
		\end{table}
		
		Next, we evaluate the impact of each modality (range-Doppler and acoustic) for drone detection and classification. First, we evaluate the acoustic only model. Here, we conduct this experiment with a one-dimensional counterpart of the proposed encoder and the drone acoustic data collected in~\cite{kolamunna2021droneprint}. As shown in Table~\ref{acoustic only model}, using the raw acoustic signal as the input and using fewer number of parameters compared to other works, we surpassed the those with a \emph{very good margin of $4$\%}. This emphasizes the effectiveness of the proposed encoder architecture. More specifically, it utilizes the raw acoustic signal without converting to Mel-spectrograms. 
		
		\begin{table}[t!]
			\caption{Comparison of the acoustic-only model for dataset in~\cite{kolamunna2021droneprint}.}
			\label{acoustic only model}
			\centering 
			\resizebox{\columnwidth}{!}{
				\begin{tabular}{lccccc}
					\hline
					\hline
					{Architecture} & {Input type} & {Parameters} & {End-to-end} & {Accuracy } & {F1-score} \\
					&& (M) && (\%) & \\
					\hline
					ResNet101~\cite{he2016deep} & Mel spectrogram & $45$ & No & $94.682$ \% & $0.95598$ \\
					AST~\cite{gong2021ast} & Mel spectrogram & $87$ & No & $85.105$ \% & $0.84783$ \\
					\textbf{Ours} & \textbf{Raw acoustic} & \textbf{$23$} & \textbf{Yes} & \textbf{$98.873$} \% & \textbf{$0.98873$} \\
					\hline
				\end{tabular}
			}
		\end{table}

		\subsection{Effect of Noise}
		
		By observing the results in Table \ref{acoustic only model}, one can argue that acoustic-only models are sufficient for drone identification. Despite the superior results in Table~\ref{acoustic only model}, acoustic is much susceptible to environmental noise. To study the effect, we conduct an experiment by varying signal-to-noise ratio (SNR). We added Gaussian noise to our captured dataset to vary the SNR. Table \ref{noise} shows that the performance is drastically reduced with decrease of the SNR. Specifically, we observe a large drop in classification accuracy when the noise is added. Last row is the dataset without noise addition. This is a deviant that, stand-alone acoustic is not suitable for robust performance. With the sensor fusion (acoustic and range-Doppler), the accuracy of both detection and classification are improved, and specifically the detection model is still robust at the low SNR levels. In particular, classification accuracy increased by $13.86$\% and detection accuracy increased by $16.3$\% at an SNR of $6$ dB. 
		\begin{table}[t!]
			\caption{Effect of noise for models with different modalities.}
			\label{noise}
			\centering 
			\resizebox{\linewidth}{!}{
				\begin{tabular}{ccccc}
					\hline
					\hline
					& \multicolumn{2}{c}{\textbf{Acoustic}} & \multicolumn{2}{c}{\textbf{Acoustic + Range-Doppler}}\\
					\textbf{SNR~(dB)} & Detection & Classification & Detection & Classification \\
					& accuracy (\%) & accuracy (\%) & accuracy (\%) & accuracy (\%)\\
					\hline
					$6$ & $77.76$~\% & $25.27$~\% & $93.56$~\% & $39.13$~\% \\
					$12$ & $87.36$~\% & $27.70$~\% & $95.87$~\% & $49.94$~\% \\
					$18$ & $91.25$~\% & $31.96$~\% & $97.21$~\% & $58.44$~\% \\
					$24$ & $96.72$~\% & $39.49$~\% & $99.87$~\% & $59.66$~\% \\
					Without noise & $99.88$~\% & $84.33$~\% & $100.00$~\% & $100.00$~\%\\
					\hline
			\end{tabular}}
		\end{table}
		\begin{table}[t!]
			\caption{Performance with different modalities.}
			\label{multi-modal}
			\centering 
			\resizebox{\columnwidth}{!}{
				\begin{tabular}{lcccc}
					\hline
					\hline
					\textbf{Modalities} & \multicolumn{2}{c}{\textbf{Detection}} & \multicolumn{2}{c}{\textbf{Classification}} \\
					& accuracy (\%) & F1-score & accuracy (\%) & F1-score\\
					\hline
					Acoustic + Range-Doppler & $100.00$~\% & $1.0000$ & $100.00$~\% & $1.0000$ \\
					Acoustic & $99.88$~\% & $0.9988$ & $84.33$~\% & $0.8597$ \\
					Range-Doppler & $76.55$~\% & $0.7654$ & $16.28$~\% & $0.0885$ \\       
					\hline
			\end{tabular}}
		\end{table}
		\subsection{Ablation Study}
		
		To further evaluate the impact of the different modalities, we compare the performance of the proposed model with different combination of modalities. The Table~\ref{multi-modal} shows the result of this comparison. For the detection task, the best performance is achieved when both the modalities (acoustic and range-Doppler) are combined, yielding perfect accuracy ($100$\%) and an F1-score of $1$ on a $15$\% test split. The standalone acoustic modality performs almost perfectly ($99.88$\% accuracy, $0.9988$ F1-score), indicating its high discriminative power in detecting drone presence. However, when using only range-Doppler, detection accuracy drops significantly to $76.55$\%, suggesting that this modality alone is less effective in detecting drones compared to the acoustic modality.
		
		For the classification task, the highest accuracy is observed when using the acoustic and range-Doppler combination, achieving $100$\% accuracy and an F1-score of $1$. The classification performance significantly deteriorates when using only range-Doppler modality, with accuracy dropping to $16.28$\%. These results highlight the critical role of acoustic features in classification, as the standalone acoustic modality achieves $84.33$\% accuracy and an F1-score of $0.8597$. The poor classification performance of the range-Doppler features suggests that this modality may be more effective when used in conjunction with acoustic data rather than in isolation.

				\section{Conclusion}
				In this letter, we investigated drone detection and classification in outdoor environments using a multi-model approach. Here, radar and acoustic sensing are used to capture drone signatures, which are fused using a transformer encoder-based deep learning model. The results show that our proposed approach outperforms the state-of-the-art approaches and our model utilizes less parameters compared to them. Also, we observed that acoustic detection is significantly affected by environmental noise, limiting its effectiveness of acoustic only approach. As future work, we expect to validate this approach for long range using a high-power radar with high sensitive directional microphones.

				\section*{Acknowledgment}
				
				We would like to extend our sincere gratitude to Sri Lanka Air Force for granting access to their drones and facilities. The computational resources used in this work were funded by the Accelerating Higher Education Expansion and Development (AHEAD) Operation; grant No. 6026-LK/8743-LK of the Ministry of Higher Education of Sri Lanka funded by the World Bank.
				
				\normalsize

				%
				%

				%
				\bibliographystyle{IEEEtran}
				\bibliography{ ref}
				%
					
					

			\end{document}